\documentclass[aps,groupedaddress,nofootinbib,11pt]{revtex4}
\usepackage[english]{babel}

\usepackage[dvips]{epsfig}
\begin{document}
%
$\quad$
\vskip .5 truecm
\title{The Inflationary 
Paradigm:
Predictions for CMB}
\vskip 1.5 truecm
\author{Renaud Parentani}
\email[]{Renaud.Parentani@th.u-psud.fr}
\affiliation{Laboratoire de Physique Th\'{e}orique, CNRS UMR 8627,
B\^atiment 210,
Universit\'{e} Paris XI, 91405 Orsay Cedex, France}
%
\maketitle
\thispagestyle{empty}
\vskip 1truecm
\centerline{{\bf Abstract }}
\vskip 0.4truecm

{
We review why the search for a causal explanation of the large scale
properties of the universe 
supports the idea that 
an extended
period 
of accelerated expansion, called inflation,
preceded primordial nucleosynthesis. 
As a consequence of inflation,
all pre-existing classical structures are washed out,
and the primordial density fluctuations
(the seeds of the large scale structures)
only result from the amplification of vacuum quantum fluctuations. 
The properties of the spectrum 
are derived and compared to 
those of the spectrum of CMB anisotropies.
The agreement is striking.


%

\par\medskip
\centerline
{\rule{2cm}{0.2mm}}\medskip\medskip\medskip



\section{Introduction}

This paper is conceived as a pedagogical introduction to
the motivations for inflation, to its mathematical settings, 
and to its implications for 
the Cosmic Microwave Background (CMB) anisotropies.\footnote{
For the other aspects of CMB physics,
we refer to the other contributions of the special
issue of {\it Comptes Rendus Physique} (2003) {\bf 4}, 8.
The full text is available on line at {\bf http://www.sciencedirect.com}.
Notice also that the present text is a slightly expanded
version of that published in the above review.}
We shall put a special emphasis on justifying the surprising result that
quantum mechanical rules must be used to compute 
the primordial spectrum.

For the non-specialist, 
two remarks should be made from the outset.
First, 
the simplest inflationary model 
has successfully passed the multiple checks
based on recent 
observational data. 
Secondly, inflation 
is a rather conservative hypothesis since it rests, 
on one hand, on the 
set of cosmological observations conventionally interpreted,
and on the other hand, on Einstein's equations, i.e., on the hypothesis 
that the action of General Relativity (GR), or a slight generalization thereof, 
governs the evolution of the 
space-time properties from cosmological scales, 
of the order of a thousand megaparsecs ($1$Mpc $ \simeq  
10^{22}$ meters), 
down to some microscopic scale which is not far from the Planck length
($l_P \simeq 
10^{-35}$ meters), 
 before the threshold of Quantum Gravity.

Following these preliminary remarks,
one should explain why the search for a causal explanation of 
the large scale properties of the universe calls for a long period
of accelerated expansion (where long is defined with respect to
the expansion rate during inflation).
This need stems from the fact that the `standard model' of 
cosmology\cite{Weinberg,Peebles},
the Hot Big Bang scenario, is {\it incomplete}: 
the causal structure from the big bang
is such that no processes
could have taken place to explain the homogeneity and the isotropy
at large scales. 
(What ``large scales'' means depends on the quantity
under consideration. For the density fluctuations on the 
last scattering surface, at the moment of decoupling, 
large scales concern distances larger than hundred 
kiloparsecs. These scales
are seen today as solid angles larger than 1 degree).
The importance of these considerations
can only be appreciated in the 
light of today's understanding of cosmology.

\section{The standard model}

The successful predictions of 
the standard model of cosmology concern {\it both} 
the evolution of spatial averaged 
quantities (temperature, densities...) 
as well as the evolution of local structures 
from density fluctuations to galaxies and clusters.
Let's see how these two aspects are woven together.
 
The CMB is extremely isotropic: 
the 
temperature fluctuations  at the decoupling time $t_{dec}$
have a relative amplitude of the order of $10^{-5}$.
When adopting GR, 
this implies that 
at that time the 
(today visible)
universe 
can be accurately described by a linearly perturbed metric. 
Indeed the primordial metric fluctuations,
defined on the top of a Robertson-Walker (RW) metric, 
are also of the order of $10^{-5}$
since they are linearly related to the 
temperature fluctuations.
We recall that the RW
metric reads
\begin{equation}
\bar{d s^2 } = - dt^2 + a^2(t) \,  d\Sigma_3^2  \, ,
\label{RW}
\end{equation}
where bar quantities describe smoothed out, spatial averaged,
quantities. The spatial part of the metric $d\Sigma_3^2$ 
describes isotropic and homogeneous 3-surfaces.
The actual metric, ignoring gravitons (the two spin 2 fluctuation
modes), can be written as
\begin{equation}
{d s^2 } = - (1+ 2 \Psi )\, dt^2 + (1 - 2 \Psi)\, a^2(t) \, d\Sigma_3^2  \, ,
\label{pRW}
\end{equation}
when the matter stress-tensor 
is isotropic, a simplifying assumption which nevertheless
captures the essential but which limits the fluctuations
to the so-called `adiabatic' modes\cite{MukhB,Langlois}.
The local field $\Psi(t, {x})$ acts as a Newtonian potential.

Since the fluctuations are $10^{-5}$, 
one can first adopt a mean field approximation
and 
consider only spatially averaged quantities.
In a second step, one can analyze the evolution 
of the fluctuations which ride on the top
of the background quantities formerly obtained.

\subsection{Mean quantities and primordial nucleosynthesis}

The evolution of the averaged metric Eq. (\ref{RW}) is 
governed  
by a single function of time, the scale-factor $a(t)$. 
The derivative of its logarithm defines $H= \partial_t{a} /{a}$
which 
enters in Hubble's law $v = H R$. 
$v$ is the relative velocity 
of comoving galaxies separated by a proper distance $R(t)$.
This distance is given by $R(t)= {a}(t) \, d$, 
where $d$ is the fixed comoving distance
between the two galaxies. This distance is defined by the
static line element $d\Sigma_3^2$ of Eq. (\ref{RW}).
In GR, $a$ obeys the
Friedmann equation, 
\begin{equation}
    H^2 =  {8 \pi G \over 3} \bar \rho + { \kappa \over   a^2 }\, .
\label{FE}
\end{equation}
The averaged matter density is denoted $ \bar   \rho$ 
and includes a possible cosmological constant\cite{PeebCC}. 
The last term in Eq. (\ref{FE}) arises from the curvature
of the homogeneous 3-surfaces.
Its today relative contribution is given by
$\Omega^{curv.}_0 = { \kappa /( H^2_0   a_0^2)}$
(where the subscript $_0$ means evaluated today).
Its value is now observationally tightly constrained:
$ \Omega^{curv.}_0  = 0.02 \pm 0.02$\cite{WMAP}.
This term is the integration constant of the 
`dynamical' Einstein equation, 
\begin{equation}
{  
\partial_t^2
 {   a }\over   a}
 = - {4 \pi G \over 3} ( \bar \rho + 3  \bar P)\, ,
\label{DE}
\end{equation}
and energy conservation 
$d \bar \rho = - 3 ( \bar \rho + \bar P) d \ln a $.
Together with the density $\bar \rho$, the pressure $ \bar P$ 
fully characterizes the matter stress tensor when imposing 
homogeneity and isotropy. 
Furthermore, when considering non-interacting components, 
$\bar \rho$ and $\bar P$ split into terms which obey 
equations of state of perfect fluids:
\begin{equation}
 \bar P_i = w_i \  \bar \rho_i \, .
\label{ESi}
\end{equation}
Dust, (cold) baryons and Cold Dark Matter (CDM) have $w =0 $, photons
and (massless) neutrinos  $w= 1/3 $, whereas the cosmological 
constant has  $w=-1$.

Energy conservation then fixes the scaling 
of each component.
For baryons, one gets $ \bar \rho_b 
\propto a^{-3}$, 
and for radiation, 
$ \bar \rho_{rad} \propto
a^{-4}$.
The 3-curvature term scales only as $  a^{-2}$ 
whereas the cosmological constant does not vary. From 
these scalings and the present values of the densities,
one immediately obtains the different stages of our
cosmological history.
In particular, the CMB temperature, $T_0 = 0.23\, {\rm meV}$, 
the effective number of massless fields\cite{Peebles},
and the relative contribution of cold matter (baryons + CDM), 
$\Omega^{cm}_0 = 0.27 \pm 0.04$ \cite{WMAP}, 
fix the redshift of the transition from radiation to matter domination:
$z_{eq} = a_0/a_{eq} -1  \simeq 3200$. 

Notice that this equilibrium occurred 
before the moment of decoupling which occurred 
when $z_{dec} \simeq a_0/a_{dec} \simeq 1100$.
Notice also that at that time, the 3-curvature contribution obeyed
$\Omega^{curv.}_{dec} <  \Omega^{curv.}_{0}/(\Omega^{cm}_0 
z_{dec}) < 10^{-4}$. 
Thus, for all processes which took place at or before decoupling,
the spatial 
metric can be taken 
Euclidean
and $\kappa/a^2$ 
in Eq. (\ref{FE}) can be dropped.
By a similar reasoning, one also concludes that 
a cosmological term or a term with $w < -1/3$ (inducing
an acceleration: $\partial^2_t a >0$) play no role
before equilibrium. 
So, unless one introduces an additional contribution
to $\bar \rho$ which dominated for a while the r.h.s of
Eq. (\ref{FE}) and then decayed, 
one arrives at the conclusion 
that the early cosmology was dominated by the radiation contribution,
$\bar \rho_{rad}$. 
We shall see in Section II.C
what are the reasons to reject this simple conclusion.

The next step consists in using 
{\it particle physics}
to inquire about the processes which took place in the early universe,
before $z_{eq}$. 
The greatest success of this standard framework
concerns the prediction of relative abundances of 
light elements by Primordial Nucleo-Synthesis\cite{Weinberg,Peebles,Nucleos}
which occurred for a redshift $z_{\rm PNS} \simeq 10^9$.
This confirms the validity of extrapolating backwards in time 
Einstein's equations in a radiation dominated universe,
at least up to redshifts of the order of 
$10^{9}$ and  temperatures of the 
order of the ${\rm MeV}$.

Moreover a detailed comparison 
of theoretical predictions 
and observations leads to a precise 
determination of the baryonic content of our universe
(through the determination of $n_b/n_\gamma$, the number 
of baryon per photon, 
and the absolute normalization of the thermal photons of the CMB).
Using the recent estimation of $H_0$, primordial nucleosynthesis
gives $\Omega_{b 0} = 0.042 \pm 0.005$,
a value in agreement with that obtained from analyzing the CMB\cite{Nucleos2}. 
So, not only the predictions of the extrapolation 
can be observationally tested, but 
independent observations agree. This
puts the standard model on extremely firm foundations.

\subsection{Fluctuations and Structures}

So far, we have ignored
the fluctuations around spatial averaged quantities. 
It is remarkable that  
the standard model of cosmology, as it stands,
also correctly predicts the evolution 
of {local structures}\cite{Weinberg,Peebles}.
These structures are described by a perturbed metric, 
as in Eq. (\ref{pRW}), and a local matter density 
$\rho = \bar \rho(t) + \delta\rho(t, {x})$. 
Their combined evolution
is governed by the {\it local} Einstein 
equations (and no longer by their restriction to RW metrics
used so far) and by matter interactions. 

Since the {primordial} fluctuations $\delta \rho/\bar \rho \simeq 
\Psi $ have an amplitude of the order of $10^{-5}$,
they evolve linearly for a long range of redshifts 
until
gravitational instabilities and radiation damping
lead to non-linear 
structures. The properties of the latter are in agreement 
with what is now observed. 
However, the best agreement is reached 
{if} one adds some Cold Dark Matter
whose density is about 10 times larger than that of 
baryons\cite{bert}.
At first sight this could be considered
as an {\it ad hoc} hypothesis to save the standard model, but 
this is no longer the case since there are now independent observations which 
are consistently interpreted as gravitational effects induced 
by CDM\cite{PeebCC}. 
The most precise estimation of
$\Omega_{{\rm CDM} 0}$ ($= \Omega^{cm}_0 - \Omega_{b0} 
= 0.23 \pm 0.04 $)
is presently extracted from 
the spectrum of CMB anisotropies. 
\subsection{The incompleteness of the Hot Big Bang scenario}

Given these two successes, it is tempting to further extrapolate 
and to inquire what happened before nucleosynthesis,
for redshifts larger than $10^{10}$ and energies larger 
than the MeV. Was there a 
baryo-genesis\cite{Riotto} which could explain 
the 
baryon-antibaryon asymmetry,
or a lepto-genesis\cite{Buch} which preceded baryo-genesis?

In this inquiry one 
searches 
processes 
by which
to explain the observed values of quantities which,
for lack of something better,
 have been hitherto treated as {initial conditions.} 
Inflation might constitute a crucial step in this endeavor.
The reason is the following.
Irrespectively of the 
high energy local 
processes which took place,
one inevitably encounters a problem
which is {\it non-local} in character:
Given the {causal} structure obtained by extrapolating 
$a(t)$, the solution of 
Eq. (\ref{DE}) driven by radiation, the
{large scale isotropy} cannot be 
explained\cite{Weinberg},\footnote{In fact
it was the reading of Weinberg's careful explanations which lead
R. Brout, F. Englert, and E. Gunzig to propose one of the first inflationary
models in 
{\it The creation of the universe as a quantum phenomenon},
Annals Phys. 115:78,1978; and {\it The causal universe},
 Gen. Rel. Grav. 10:1-6,1979.}
as can be seen from Figure 1.

\begin{figure}[ht]
\epsfxsize=7cm
\centerline{{\epsfbox{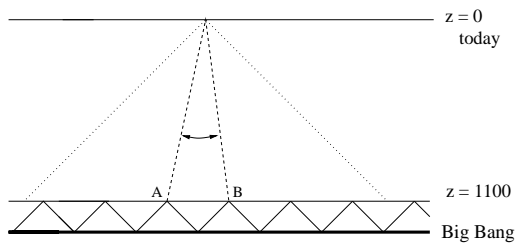}}}
\caption{ {\it The causal structure from the Hot Big Bang.}
In a radiation dominated universe, there is a {\it space-like}
singularity at $a=0$ which is situated
at a {\it finite} conformal time $\eta = \int_{t_{B\!B}}^{t} dt/a$.
Hence forward light cones from the big bang 
have, at time $t$, a finite (proper) radius
equal to $a(t) \eta(t)$ which defines the 
so-called {\it particle horizon}.
The proper distance between the points $A$ and $B$
is equal to $a\Delta\eta$ evaluated on the last scattering surface ({\it lss}).
Hence the matter systems at $A$ and $B$ 
have never been in contact. Thus no 
(causal) process could possibly explain why the temperature
fluctuation between 
$A$ and $B$ 
is only of the order of $10^{-5}$. 
Moreover 
the today visible universe (in dotted lines)
encompasses 
about $5.\, 10^{4}$ disconnected patches on the {\it lss}
with fluctuations within that range.}
\end{figure}

There is an interesting and complementary way to consider this
incompleteness. It concerns the {origin} 
and the evolution of {\it large scale anisotropies}. 
Since the fluctuation modes propagate 
in RW metric,
their comoving wave vectors $ k$ are conserved.
Hence, when treated linearly, the Fourier modes
$\Psi_k = \int\!d^3x \, e^{ikx} \Psi$ and
$\delta \rho_k = \int\!d^3x \, e^{ikx} \delta \rho$
evolve independently of modes
with $k' \neq k$.
The important point is that this evolution crucially depends on the
ratio $\lambda/R_H$ where $\lambda=a(t)/k$ is the physical wave length 
and where $R_H= 1/H(t)$ is the Hubble radius ($c=1$).
When $\lambda/R < 1$,
the modes are `inside' the Hubble radius, and they oscillate.
On the contrary, on super-horizon scales,\footnote{In the Hot Big Bang
cosmology, the Hubble horizon $R_H(t)$, which is locally defined, 
coincides with the globally defined particle horizon $a(t) \eta(t)$
where $\eta$ is the conformal time evaluated from the singularity.
This equality no longer applies in inflation,
thereby offering a solution to the causality problem 
(sometimes also called the {\it horizon problem}) 
we are discussing.}
radiation dominated universe i.e.
when their wave length extends beyond the Hubble radius,
modes are frozen (or decay).
Because $k$ is conserved 
 it is appropriate to follow the evolution of $\lambda/R_H$
 by writing it as $k^{-1}/d_H$ where
\begin{equation}
  d_H(t) = {  R_H(t) \over   a(t)} = {1 \over \partial_t{   a}} \, ,
\label{dH}
\end{equation}
gives the comoving size of the Hubble radius at time $t$.

These considerations become crucial when questioning
the {\it origin} of primordial fluctuations on super-horizon 
scales (we know they exist since they determine the CMB anisotropies
at large scales, for solid angles larger than a degree). 
Indeed since $\partial_t^2 a < 0$ in a radiation dominated universe,
as seen from Eq. (\ref{dH})
and Eq. (\ref{DE}),
$ d_H$ always increased.
Therefore the modes which were still outside 
$d_H$ on the last scattering surface 
have been frozen since the big bang. 
Hence
their amplitudes 
cannot result from processes and 
can thus only be fixed from the outset
by {\it initial conditions}.

\section{Inflation}

Inflation
is 
the price to pay to reject this vexing outcome.\footnote{
Topological defects resulting
from a phase transition could have been another
possibility.
However, 
recently 
obtained properties of 
CMB anisotropies (i.e. the temporal coherence of the modes
revealed by the multiple 
``acoustic'' peaks,
see later in the text)
rule out 
this possibility\cite{PPeter,FirstY}.}
From the above considerations, a 
{necessary} condition
for allowing physical processes to have taken place is 
that 
there 
was a time when $d_H$ was {\it larger} than today's Hubble
scale $d_{H0}$. Therefore, between then and the 
beginning of the radiation era,
$d_H$ must have tremendously {\it decreased}. Thus from 
Eq. (\ref{dH}) there must have been a ``long'' period of accelerated expansion,
see
Figure 2. 

\begin{figure}[ht]
\epsfxsize=7cm
\centerline{{\epsfbox{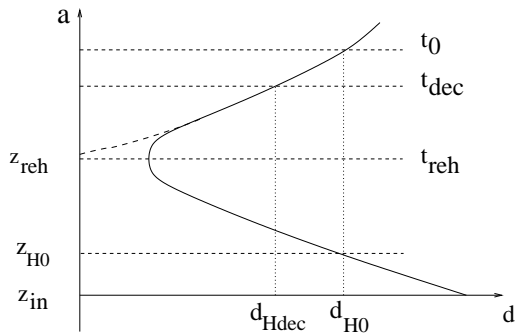}}}
\caption{
{\it The evolution of the comoving value of the Hubble radius $d_H$.}
A vertical line is a fixed comoving scale. The symbols
$d_{H0}$ and $d_{Hdec}$ designate the Hubble scale 
evaluated today 
and at the decoupling respectively. 
The evolution of $d_H$ with or without inflation
splits near some high redshift $z_{reh}$. 
The Hot Big Bang branch is indicated by the dashed line
whereas the inflationary branch is marked by the continuous line. 
$z_{in}$, $z_{H0}$,  and  
$z_{reh}$
give the respective values of the redshift when inflation starts,
when the scale $d_{H0}$ exits $d_H(t)$, and at {\it reheating} when inflation stops.
For inflation to be successful (i.e., to remove
the need of fine tuning the initial conditions of fluctuation modes) 
one must have $d_H(t_{in}) \gg d_{H0}$, as we shall explain in Section III.A.} 
\end{figure}

At this point, inflation is simply a kinematic hypothesis 
which allows processes to have taken place.
(Notice however that the dilution of scales associated with the 
{decrease} of $d_H$ during inflation
solves other problems of cosmology\cite{Guth,Linde}).
In order to proceed, several issues should be confronted. 
One should first dynamically realize inflation, i.e., 
find an ``engine'' which could be responsible for it. 
Secondly one should 
identify the mechanism giving rise to the primordial spectrum\cite{Mukh}. 
The third issue concerns the {\it reheating}\cite{Lindebook,reheat}: 
the collective process which damps the expansion
and liberates heat
thereby ending the period of inflation 
and starting
the radiation dominated period
which is necessary for 
primordial nucleosynthesis to take place.

It should be noticed that to a large extend these three issues 
could be addressed separately. In particular, it turns out that the
mechanism giving rise to the spectrum operates quite irrespectively
of the particular dynamical realization of inflation one adopts.
Hence, 
different dynamical models could end up with very similar spectra.
Therefore, in view of our interest in the CMB,
we shall focus on this mechanism and the properties of 
the spectrum it produces.
We shall also restrict ourselves to the simplest inflationary model,
that of a single massive scalar field. For a large
panorama of inflationary models we refer to \cite{LiLy}.

\subsection{The inflationary background}

To have {\it successful} inflation requires two things.
On one hand, the accelerated expansion should last
enough that
today's Hubble scale
$d_{H0}$ be  inside the Hubble radius 
at the beginning of inflation, see Figure 2.
To characterize the ``duration'' of inflation it is convenient to
introduce the parameters 
$N_{min}=\ln(a_{reh}/a_{H0})$
and $N_{tot} = \ln(a_{reh}/a_{in})$
which respectively give the minimum number of {\it e-folds}
for inflation to be successful and the total number 
of {e-folds} from the beginning of inflation 
to its end at the reheating. 
On the other hand, even though inflation needs not to be homogeneous,
the patch which inflates
should be sufficiently large and homogeneous that the gradients
be negligible up to a scale larger than 
today's Hubble scale $d_{H0}$. If one of these conditions is not met, 
inflation would fail to explain the isotropy of the CMB
up to that scale.

To have accelerated expansion
in GR requires $\bar \rho + 3 \bar P < 0$,
see Eq. (\ref{DE}).
To have inflation thus requires that this condition
be satisfied for a long time but only in the above mentioned patch.
This can be fulfilled by {introducing} a scalar field,
named {\it inflaton}, 
which 
possesses
an expectation value $\bar \phi$ which 
obeys three conditions: 
$\bar \phi$ must 
dominate {all} other contributions to $\rho$ and $P$,
it must slowly decay, and it must be sufficiently homogeneous.
When these are met, one can write the inflaton as
\begin{equation}
\phi(t,x) = \bar \phi(t) + \delta\phi(t, {x}) \,  ,
\label{inflon}
\end{equation}
and neglect non-linear terms in the fluctuations $\delta\phi, \Psi$
since $ \Psi \propto \delta\phi/\bar \phi \ll 1$.
Then,  in the patch, $\phi$ 
obeys the scalar field equation 
in a RW metric:
\begin{equation}
\partial_t^2 \phi + 3 H \partial_t \phi - 
{ \Delta  \phi \over a^2} + m^2  \phi = 0 \, ,
\label{eqinflon}
\end{equation}
where $\Delta $ is the Laplacian associated with the static line
element $d\Sigma_3^2$ of Eq. (\ref{pRW}).
From Eq. (\ref{eqinflon})
one finds that the slow decay of 
$\bar \phi$ follows
from $3 H \partial_t{\bar \phi} \simeq  - m^2 \bar \phi$, 
i.e., $\bar \phi$ is dragged to zero at a rate given by $m^2/3H$.

We are now in position to verify that $ \bar \phi $ leads to 
inflation when the above  conditions 
are satisfied.
The energy density
and the pressure 
are dominated by the contribution of $\bar \phi$ given by 
$\bar \rho_\phi, \bar P_\phi = 
(\partial_t \bar \phi)^2 /2 \pm m^2 \bar\phi ^2/2$. 
When the decay is slow enough,  
the kinetic term is negligible. In this case
one gets $\bar P_\phi / \bar \rho_\phi \simeq -1 $
and hence accelerated expansion, see Eq. (\ref{DE}).
As long as the mass term is larger than the kinetic term, 
$ \bar \phi $ thus acts as a (slowly decaying) cosmological constant. 
To characterize this decay, it is appropriate to introduce 
the {\it slow-roll} parameter 
\begin{equation}
\epsilon= - \frac{\partial_t H}{H^2} = -
\frac{d\ln H }{d \ln a } \ll 1 \, .
\end{equation}
Algebra then gives $m^2/3H = \epsilon H$ for the decay rate,
and
$\bar P_\phi /\bar \rho_\phi = -1 + 2 \epsilon/3 $
for the equation of state.  
One also finds that 
$N_{tot} = 1/2\epsilon = 2 \pi G \, \bar \phi_{in}^2$, thereby relating 
the total number of e-folds to $\epsilon$ and to the initial value
of the inflaton in Planck units.

It remains to make contact with (micro)physics. It is generally 
believed that the reheating process has something to do
with Grand Unification Theories~\cite{Guth,Lindebook}.
If this is correct $T_{reh}$, the reheating
temperature, should be close to GUT scale, near $10^{14} {\rm GeV}$. 
In this case, there must be at least 60 e-folds of inflation.
This condition follows from:
$N_{tot}
> N_{min} 
\simeq \ln(d_{H0}/d_{Hreh})
\simeq \ln(a_0/a_{reh}) 
\simeq \ln (T_{reh}/T_0)
\simeq 60$.
(The first two $\simeq$
follow from
scaling laws:
$d_H \propto 1/a$ during inflation and $d_H
\propto a$ during the radiation era.)

Before considering the fluctuations of the inflaton, 
it should be stressed  that
the initial value of $\bar \phi$ obeys 
$\bar \phi_{in} > N_{min}^{1/2} M_{P}$, 
where $M_P= G^{1/2}$ is the Planck mass when $c=\hbar =1$.
That is, the slow-roll conditions send us inevitably 
{\it above} the Planck 
scale. However there is no particular reason to 
believe that the settings we are using make sense in this regime.
More complicated inflationary models based on two fields do 
not suffer from this disease\cite{LiLy}.
Nevertheless the question of the nature of the inflaton is still open. 
There are basically two attitudes.
The dominant attitude\cite{Guth,Lindebook}
is that inflation belongs to particle
physics and occurs sufficiently below the Planck scale so that gravity 
can be safely treated by classical GR.
This  
option raises a very difficult question\cite{Brandy} related to the 
cosmological constant problem: If gravity can all the way be 
treated classically, why the mechanism which screens the
vacuum energy during the whole radiation era (where several 
transition should have taken place thereby giving rise to 
non-vanishing vacuum energy) did not screen as
well the potential energy of the inflaton during inflation?
One should therefore 
not exclude the alternative possibility 
that the inflaton be merely a 
phenomenology which aptly characterizes,
as e.g. in $R^2$-inflation\cite{Staro},
the background and the 
fluctuations in a domain wherein quantum gravity 
(or stringy effects) could still play an important role.
As far as the mechanism giving rise to
the primordial spectrum is concerned, 
these alternatives are equivalent\cite{Mukh}. 



\subsection{The primordial spectrum}

To identify this mechanism,
we need to consider the 
combined evolution of
the inflaton and the metric perturbation, $\delta \phi$ and $\Psi$.
Assuming we can 
work to first order in these fluctuations,
GR gives us a set of equations for their Fourier components,
$\delta \phi_k$ and $\Psi_k$,
see \cite{MukhB} for details. 
Two important points should be mentioned.
First the $0i$ Einstein equation gives 
\begin{equation}
{1 \over a}\partial_t (a \Psi_k) = 
{4 \pi G }  \partial_t \bar \phi  \, \delta\phi_k \, .
\label{0ieq}
\end{equation}
From this constraint equation one learns that $\Psi_k$ decays like $1/a$
in the absence of matter density fluctuations. 
Hence the primordial metric fluctuations
are sustained by matter fluctuations.
From the linearization procedure, one also learns that, 
during inflation, the {\it dominant} matter fluctuations are
those of the inflaton because the background energy
$\bar \rho$ is dominated by $\bar \phi$. (When considering inflationary models
with several scalar fields, 
a particular combination of their fluctuations, named {\it adiabatic},
drives the metric fluctuations $\Psi_k$. The other fluctuations, 
called iso-curvature, could nevertheless play important roles\cite{Langlois}.)

The second important fact is that the modes $v_k$, defined by
$v_k /a = \delta \phi_k +   \Psi_k (\partial_t \bar \phi/ H )$, 
are those of a canonical (and gauge invariant) field. 
Since all fluctuations obey linear equations 
it is indeed crucial to identify the combination
which forms a canonical field  
because its {\it normalized} quantum mechanical
fluctuations will be taken into account in the sequel.
The modes $v_k$ behave like
 harmonic oscillators with time dependent frequency:
\begin{equation}
\partial_\eta^2 v_k + (k^2 - \xi) v_k = 0 \, ,
\label{Meq}
\end{equation}
where $\eta$ is the conformal time ($d\eta = dt/a$) 
and where $\xi$ is a function of the background solution
approximatively given, in the slow-roll limit, by
$2/\eta^2$. Hence, in the early past, for $\eta \to -\infty$, Eq. (\ref{Meq}) 
reduces to a standard harmonic oscillator. However when $k^2 = {\xi}$, 
which corresponds to $k \simeq 1/d_H(t)$, 
the mode exits the Hubble radius (also often called Hubble horizon) 
and stops oscillating.
One gets a growing ($\propto a$) and decaying mode
($\propto 1/a)$. Thus $v_k/a$ becomes constant determined
by the growing mode. To evaluate the norm of this constant 
one needs an initial condition for $v_k$. 
In preparation of the subsequent analysis,
let us impose that $v_k$ has a unit Wronskian
and contains only positive frequencies
for large values of $kd_H(t_{in})$, i.e., 
when the mode is well inside the Hubble radius.\footnote{In 
the present classical treatment, this choice is purely mathematical. 
It will become physically meaningful when we shall
use second quantized rules in the next subsection.
Let us also point out that we shall consider only the norm 
of $v_k/a$ since we have in mind that the fluctuations should
be described by a statistical ensemble which fixes only the
power of the spectrum.} 
After horizon exit, one finds that the {\it frozen} value of $v_k/a $ is  
\begin{equation}
\vert (v_k/a)^{fr} \vert^2 = {H_k^2 \over 2 k^3} = 
\vert \delta \phi_k^{fr} \vert^2 \, \left(1+ O\!\left(\epsilon\right)\right)\, ,
\label{Fsp}
\end{equation}
where $H_k$ is the value of $H(t)$ at horizon exit,
when $k=1/d_H(t)$. 

Then Eq. (\ref{0ieq}) implies that 
$\Psi_k$ also tends to a constant after horizon exit.
The frozen value is given by
$\Psi_k^{fr} \simeq - \delta\phi_k^{fr}/\bar \phi $,
i.e. $\vert \Psi_k^{fr} \vert^2 = 4\pi G\, \epsilon \vert \delta\phi_k^{fr}\vert^2$.
However,
there is still a subtlety:
$\Psi_k^{fr}$ gets multiplied by 
$2/3\epsilon$ when inflation stops. 
This follows from the existence
of a conserved quantity\cite{MukhB} which is
proportional to $\Psi_k (w+5/3)/(w+ 1)$, where $w=\bar P/\bar \rho$.
Hence when the equation of state changes
from inflation 
to radiation,  
the amplification factor is $2/3 \epsilon$.

In brief, from the combined evolution of 
the background variables $a, \, \bar \phi$
and the modes $\Psi_k, \, \delta \phi_k$, inflation tells us that
in the radiation dominated era, before horizon re-entry, the {\it primordial}
spectrum of the metric fluctuations $\Psi_k$  is given by
\begin{equation}
\vert \Psi_k^{prim} \vert^2 = {4 \over 9}  {4\pi G \over \epsilon_k } \vert (v_k/a)^{fr} \vert^2  \, .
\label{prS}
\end{equation}
We have added a subscript $k$ to  $\epsilon$ because, in general,
it (slowly) varies with time and hence on $k$ through $k/a=H$.
The primordial spectrum 
it thus given in terms of the background equation
of state $\bar P/\bar \rho$ and the norm of the frozen value of $v_k/a$
evaluated during inflation, just after horizon exit. 
This is the first non-trivial outcome of inflation.
From Eq. (\ref{Meq}), the r.h.s. of Eq. (\ref{prS}) 
can be computed {\it if} we know 
the amplitude
of $v_k$ at the onset of inflation.
Therefore, so far,
we have only ``postponed'' 
the problem of the initial conditions, from some high redshift 
before nucleosynthesis, to some still higher redshift before 60 e-folds
of inflation.

\subsection{Initial conditions and QFT}

It thus remains to confront the 
question of the initial conditions of the inflaton fluctuations.
This question only concerns 
the values of $k$ which correspond to scales which
are today visible. We shall call them the {\it relevant} scales.
When inquiring about their initial conditions,
one encounters a major surprise: {\it one must abandon classical settings.}
The reason is the following. 
When a mode $v_k$ is well inside the Hubble horizon, 
it behaves as 
a massless mode, see Eq. (\ref{Meq}).
Hence,
its energy density behaves like radiation
and scales like $1/a^4$. 
Thus, {if} $v_k$ were to possess a classical amplitude inside the Hubble horizon,
one would reach an inconsistency because when propagating  
backwards in time, 
its energy density would, after few e-folds, overtake
the background density $\bar \rho_{\phi}$ 
which is almost constant. 
But this would violate our assumption that there {\it is} inflation
which implies that $\bar \rho_{\phi}$ {is} the dominant contribution.
So, unless one fine tunes 
the number of {\it extra} e-folds, $N_{extra}= N_{tot} - N_{min}$, 
the relevant modes cannot have
classical amplitudes at the onset of inflation.\footnote{For the 
same reason, the curvature relative contribution 
should be exponentially small when evaluated today:
$\Omega_0^{curv.} \ll 1$. In other words, in inflation, 
the {\it relevant} part of the spatial sections is flat.}

One must therefore look for a quantum mechanical origin
of the primordial spectrum.\footnote{ 
If it turns out that the {\it large scale structures} of our universe 
are of quantum origin, 
this would constitute the triumph of quantum mechanics 
which was elaborated from 
atomic and molecular spectra.}
When adopting the settings of quantum field
theory in curved space\cite{BDavies}, 
one should re-address the question of the initial
conditions. It is now formulated  
in terms of the occupation number $n_k$ 
of the quanta of the field operator $\hat v$ in  Fock space. 
For relevant $k$,
one finds that $n_k$ must vanish at the onset of inflation.
Indeed, the energy density carried by these quanta 
($\propto n_k k^4/a^4$) also scales like $a^{-4}$ and
would therefore violate, as in classical settings, the hypothesis that
the background energy density is dominated by $\bar \rho_{\phi}$.
So one is left with the vacuum as the unique possibility.
 It is then remarkable that the vacuum energy contribution evades
a potential inconsistency which would have otherwise ruined inflation:
Because of the subtraction of the zero-point energy\cite{BDavies},
the vacuum energy does {\it not} grow like $1/a^4$ and stays
much smaller than the background density $\bar \rho_{\phi}$. 
Hence vacuum is the only initial state of relevant modes 
which is consistent with inflation.
Notice that nothing can be said about the (irrelevant) infra-red 
modes 
with $k \ll 1/d_{H0}$
because their energy density does not diverge at the onset of inflation. 
Notice also that  nothing should be said about them
because they are viewed today as part of the homogeneous background
and cannot be identified as modes. 

There is a complementary way to understand how inflation
answers the question of the initial conditions.
Instead of asking what happens when propagating modes
backwards in time, lets directly consider what could possibly be the initial state
at the onset of inflation. 
In the homogeneous patch, there is an energy scale: $H$.
Hence, the occupation number of quanta of (proper) frequency 
$\omega_{in} =k/a(t_{in})$ 
must obey a kind of Wien law 
$n(\omega_{in}) \propto e^{-\omega_{in}/H}$ for $\omega_{in} \gg H$.
On the other hand, at the onset of inflation,
the initial frequencies of all relevant modes
obey $\omega^{\rm relevant}_{in} > H z_{in}/z_{H0}$, where $z_{H0}$ is the redshift when 
the largest visible scale 
$d_{H0}$ exits the Hubble radius, see Figure 2. 
Therefore, when the number of {extra} e-folds, 
$N_{extra}= N_{tot} - N_{min}$, 
is larger than, say $5$, one finds that 
all relevant modes {\it are} in their ground state.
Concomitantly to this vacuum condition, at the onset of inflation, 
the proper size of scale $d_{H0}$ 
was given by $H^{-1} z_{H0}/z_{in }=  H^{-1} e^{-N_{extra}}$. 
So unless one constraints $N_{extra}$ to be smaller than $\ln(M_{Planck}/H)
\simeq 13$, our universe was initially inside a Planck cell.
To appreciate the unavoidable character of this
conclusion, we invite the reader to draw the ``trajectory'' of the
(proper) Planck  length $l_P$ in ($a,d$) plane of Figure 2.


So, in inflation, the choice of the initial state does not follow from 
a principle (e.g. some symmetry) but it is fixed by the 
kinematics of inflation itself.
This stems from the blue-shift effect encountered in a
backward in time propagation which sends the proper
frequencies $\omega = k/a$ of relevant modes way above $H$. 
At this point it is interesting to mention the
precise analogy between these aspects and 
black hole physics. 

Classically black holes cannot radiate and this is 
deeply connected to the {\it no hair} theorem:
stationary solutions are characterized only
by mass, angular momentum and electric charge because 
all multipoles are radiated away\cite{MTW} after a few {e-folds},
where the unit of time is here given by the inverse of
the ``surface gravity'' of the hole. 
(For a neutral spherically symmetric hole, this characteristic
time is the Schwarzschild radius.)
So if after some e-folds some radiation is emitted by a black hole,
it must be of quantum origin, i.e. it must be Hawking radiation\cite{Hawk,GO}.
Similarly, for the relevant scales
in inflationary cosmology, 
the geometry would classically be bald,  no hair;
because, when $N_{extra}=N_{tot} - N_{min} \gg 1$,
all pre-existing classical structures are so much
diluted that today they would still be part of the homogeneous background. 
So if some structure is found at smaller scales, 
it must be of quantum origin. 
In addition, in both cases, the properties of the spectrum is 
fully determined by the geometry because the initial state 
of relevant modes is the ground state.\footnote{It should also
be noticed that this conclusion 
has recently been 
questioned both in black hole physics
and in inflationary cosmology
under the appellation {\it the trans-Planckian question}, for a review 
see \cite{tp}.
The question arises from the fact that
quantum fields in a curved space-time
should be a valid approximation of quantum gravity 
only below the Planck scale.
However, as we saw, the initial state is specified for
frequencies well above that scale. Hence
we should worry about the validity of 
the predictions we obtained. 
To confront this question, a phenomenological approach
proposed by Unruh\cite{unruh} has been thoroughly  investigated. 
The (disputed) conclusion\cite{bhtp,ictp} is that modifications 
of the standard results are unlikely because the properties of the
spectrum are generically {\it robust} even when 
drastically modifying the physics above the Planck scale.}

\subsection{Predictions and observational data}

Having understood how inflation fixes 
the initial condition of fluctuations, we can bring together the 
various results. Since $v(t,x)$ is a canonical field, the positive 
frequency modes $v_k(t)$ should have unit Wronskians in order for 
the creation and destruction operators to obey the canonical commutation
rules. Therefore $<\hat v_k \hat v_{k}^\dagger>$, 
the v.e.v. of the Fourier transform of the field operator,
is equal to $1/2k$ ($\hbar = 1$)
at the onset of inflation. 
Then Eq. (\ref{Fsp}) gives the v.e.v. after horizon exit.
(In quantum optical terms, one would say that the freezing out of the 
modes at horizon exit induces a
parametric amplification of vacuum fluctuations
which leads to extremely squeezed {two}-mode states\cite{Grishuk}. 
When considering the modes near horizon re-entry, 
the high squeezing manifests itself by the 
{\it temporal coherence} of the modes, a 
property specific to inflation, see e.g. \cite{Albrecht} and below.)
Using Eq. (\ref{prS}), the primordial power spectrum $P^{prim}_k$,
defined by
the two-point function
\begin{equation}
\langle \Psi(t, x+y) \, \Psi(t, y) \rangle 
= {1 \over 2 \pi^2} \int_0^\infty {dk \over k}{\sin(kx) \over kx} \, P^{prim}_k\, ,
\label{Psp}
\end{equation}
is given by
\begin{equation}
P^{prim}_k= k^3 \vert \Psi_k^{prim} \vert^2 = { 8 \pi G \over 9}
{H^2_k \over \epsilon_k} \, .
\label{Posp}
\end{equation}
In Eq. (\ref{Psp}), $t$ should be in the radiation dominated era,
well before horizon re-entry. When getting close to horizon re-entry,
$\Psi_k$ starts re-oscillating.

In brief,
the simple model of inflation on a single scalar field
{\it predicts} that the primordial spectrum 
should enjoy the following properties:
\begin{itemize}
\item
The power spectrum Eq. (\ref{Posp})
is nearly scale invariant.  This 
results from the stationarity
of the process: parametric amplification of vacuum fluctuations, $k$ after $k$.
In fact the $k$-dependence of $P_k^{prim}$
only originates from the slow 
evolution of the background with $t$
evaluated at horizon exit, when $k=1/d_H(t)$. 
\item
The fluctuation spectrum forms a Gaussian {\it ensemble} --
because the vacuum is a Gaussian state,
and because relevant modes were initially in their ground state. 
This means that the {\it probability} to find $\Psi_k^{prim}$ 
with a given amplitude is Gaussian, and that the l.h.s. of
Eq. (\ref{Psp}) should be 
interpreted as
an {ensemble} average.
(Non-Gaussianities are not expected to develop in the early universe
since the rms fluctuations are $10^{-5}$\cite{malda}.)
\item
The modes $\Psi_k$ are {coherent} in time
at horizon re-entry. By coherent one means the following.
When $k/a\simeq H$, each $\Psi_k$ starts to 
re-oscillate. Hence the general solution is governed by {two} 
independent quantities.
Inflation predicts that
the combination representing
the decaying mode vanishes, 
because it has had all the time to do so.  
\end{itemize}
Observational data tell us that the spectrum does enjoy the following properties:
\begin{itemize}
\item
The power spectrum is nearly scale invariant:
It has a rms amplitude $\simeq 10^{-5}$ (which fixes $H_k 
\simeq 10^{-6} \, M_P$) and it is parameterized
by a spectral index, defined by $n_k = d\ln P/d \ln k$,
which obeys\cite{WMAP} 
$n_k = 0.93 \pm 0.04$ and $dn_k/d\ln k = -0.03 \pm .02$ at a scale
which represents $0.5${\%} of $d_{H\, 0}$, the conformal
scale of 
today's Hubble radius.
\item
Non-Gaussianities have not been found, and
the spread of the data 
(the {\it cosmic variance}) is compatible
with 
standard deviations given the finite number of independent observables. 
\item
Narrow ``acoustic'' peaks are observed in the temperature fluctuations spectrum.
These are by-products of the temporal coherence of the $\Psi_k$ at horizon re-entry. 
Moreover, when considering the {\it polarization}
of the CMB, one finds an anti-correlation peak which is 
``the distinct signature of primordial adiabatic fluctuations''\cite{FirstY}.
This peak excludes that cosmic strings or textures could be a relevant
mechanism for primordial fluctuations.
\end{itemize}

In addition to these three points, the properties of the
CMB anisotropy spectrum tell us that 
the contribution of the spatial curvature
obeys $\Omega_0^{curv.}= 0.02 \pm 0.02$ in agreement with the
prediction that it
should vanish,
see footnote 5.
In brief, the 
outcome of the above comparison is that {\it the generic predictions of
the {\rm simplest} inflationary model are in accord
with observational data.}

\subsection{Perspectives}

It should be clear to the reader that 
several aspects  have not been discussed
in the present text.
These include polarization,
galaxy spectra (to have access to the power spectrum at 
smaller scales) as well as gravitational waves and iso-curvature modes\cite{JHu}. 
By taking all these  into account,
a finer understanding of the predictions of inflation
will be obtainable and one might envisage 
constraining  inflationary models
by observational data (already available and also to come).
To give a flavour of this 
program, let us briefly present
some of the points under current
investigation. They concern {\it refinements} of the simple model
we presented.
\begin{itemize}
\item
The primordial power spectrum can be {\it phenomenologically} expressed in terms of
{several} {slow-roll} parameters which characterize the inflaton action and/or
the evolution of the background\cite{Lid}.
It is hoped that these parameters, and therefore the inflaton action
(or its potential), could be determined from observational data.
In particular, degeneracies amongst inflationary models
could be removed by the detailed properties of the polarization spectra.
\item
Primordial non-Gaussianities can be obtained when
iso-curvature modes are considered\cite{Uz}. 
Theoretical studies might orient and ease observational search.
\item
Inflation also predicts that gravitational waves should be produced.
Their spectrum is given by $P^{grav}_k= 4\pi G  \, H_k^2$,
in the place of Eq. (\ref{Posp}).
Being characterized by 
different spectral indices, the observation of both 
$P^{grav}_k$ and $P^{prim}_k$
would allow
to perform consistency checks which 
could confirm (or rule out) inflation.
(Notice however that the lower value of $P^{grav}_k$
might delay (or even exclude) the detection of primordial
gravitational waves.)
\end{itemize}

Besides these phenomenological questions, 
deep fundamental questions persist, in 
particular: Who is the inflaton? What is the re-heating temperature?



\vskip .7 truecm 

{\bf Acknowledgements}
I am grateful to N. Obadia and D. Campo for their comments and advice during the
writing of this paper. I am also indebted to P. Peter and J-Ph. Uzan 
for useful remarks.}
\vskip .7 truecm 


\end{document}